\documentstyle[12pt,epsfig]{article}

\begin{document}
\begin {center}
{\bf {\Large
Shadowing in the low energy photonuclear reaction
} }
\end {center}
\begin {center}
Swapan Das \footnote {email: swapand@barc.gov.in} \\
{\it Nuclear Physics Division,
Bhabha Atomic Research Centre,  \\
Trombay, Mumbai-400085, India \\
Homi Bhabha National Institute, Anushakti Nagar,
Mumbai-400094, India }
\end {center}

\begin {abstract}
The photonuclear reaction in the multi-GeV region occurs because of the
electromagnetic and hadronic interaction. The later originates due to the
hadronic fluctuation, i.e., vector meson, of the photon. The total cross
section of the reaction is shadowed because of the vector meson nucleus
(hadronic) interaction.
To
estimate it quantitatively, the cross section of the photonuclear reaction
was calculated in the low energy region ($\sim 1-3$ GeV) using simple vector
meson dominance (SVMD) model, i.e., the low lying vector mesons ($\rho^0$,
$\omega$ and $\phi$ mesons) were considered.
The
nuclear shadowing is reinvestigated using generalized vector meson (GVMD)
model, where the higher $\rho$ meson effective state ($\rho^\prime$ meson)
is taken into account along with the low lying vector mesons.
Using 
GVMD model, the shadowing and the total cross section of the photonuclear
reaction are calculated in the above mentioned energy region. The calculated
results have been compared with the measured spectra.
\end {abstract}


PACS number(s): 25.20.-x, 25.20.Dc

\section{Introduction}

The ratio of the total cross section of the nuclear reaction to $A$
(nuclear mass number) times of that of the nucleonic reaction is called
transparency $T$ of the nuclear reaction.
The
transparency of the photonuclear reaction, i.e., 
$T = \frac{ \sigma_t^{\gamma A} }{ A\sigma_t^{\gamma N} }$, in the GeV
region is found less than unity. This phenomena is called shadowing in the
photonuclear reaction.
Indeed,
it is the feature of the hadron nucleus reactions. Therefore, the shadowing
in the photonuclear reaction indicates the hadronic behavior of photon
\cite{bauer, yennie, donnachie}.
The
photon can fluctuate into hadronic states, i.e., vector mesons, which
undergo multiple scattering while traversing through the nucleus. The
quoted scattering leads to the reduction of the cross section of the
photonuclear reaction.

\begin{figure}[h]
\begin{center}
\centerline {\vbox {
\psfig{figure=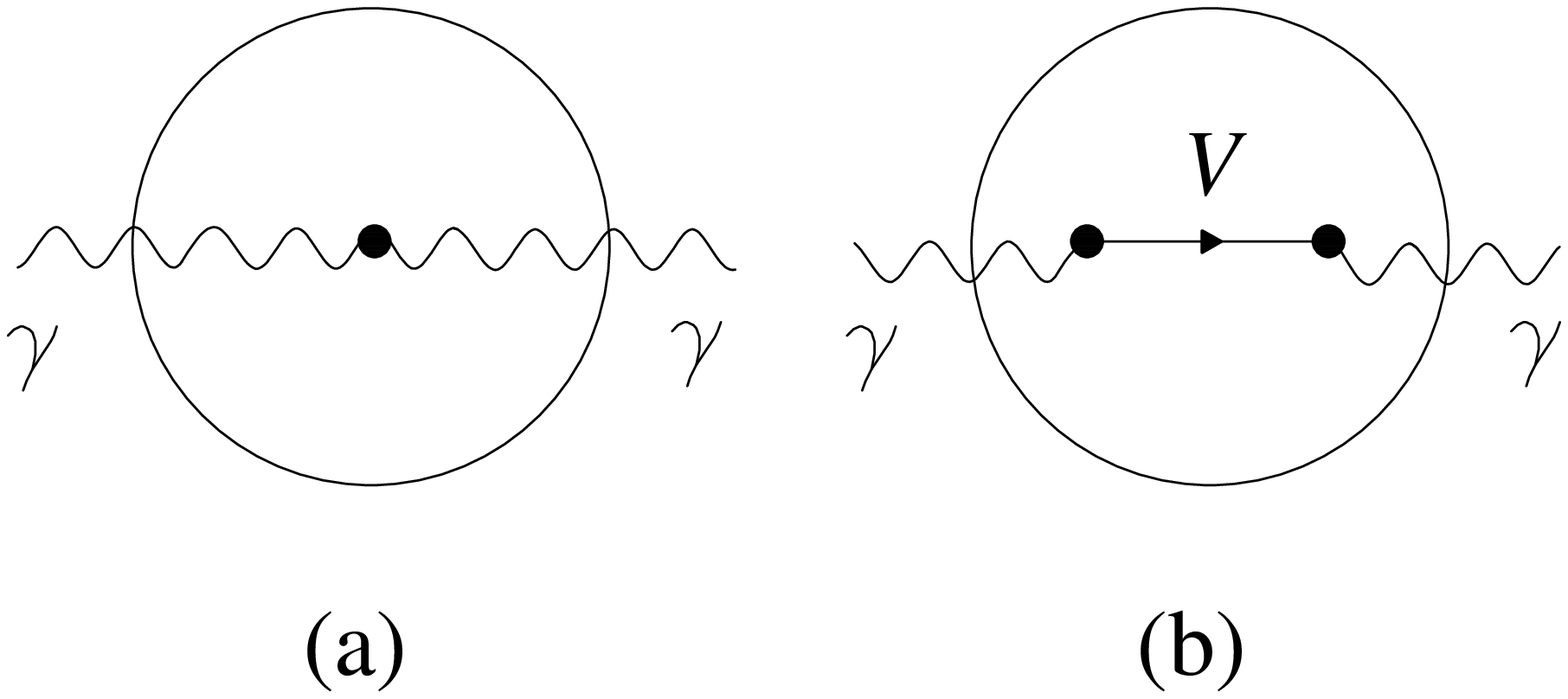,height=05.0 cm,width=10.0 cm}
}}
\caption{
The pictorial presentation of Compton scattering on the nucleus.
}
\label{FgVp}
\end{center}
\end{figure}

The total cross section of the photonuclear reaction can be evaluated by
applying the optical theorem on the amplitude of the forward Compton
scattering illustrated in Fig~\ref{FgVp}.
The
forward scattering of photon from single nucleon, shown in
Fig.~\ref{FgVp}(a), occurs due to the electromagnetic interaction, i.e., it
does not indicate the hadronic content of photon. Therefore, it leads to
the unshadowed cross section of the reaction.
Fig.~\ref{FgVp}(b)
describes the vector meson mediated forward Compton scattering on a
nucleon in the nucleus, which gives rise to the nuclear shadowing because
of the hadronic interaction of the vector meson with the nucleus.
Using
Glauber model, the amplitude of the forward nuclear Compton scattering can
be expressed in terms of the vector meson nucleon interaction
\cite{yennie}.
In fact,
the shadowing in the high energy (multi-GeV) photonuclear reactions can be
understood quantitatively using Glauber formalism \cite{bauer, donnachie}.

The data of the photonuclear reaction at $E_\gamma = 1 -2.6 $ GeV
\cite{falter, bianchi} shows early onset of the nuclear shadowing which
could not be  explained by newer models \cite{pillar, boffi}, i.e., the
calculated results, according to these models, either slightly
underestimate or overestimate the data.
Falter
et al., \cite{falter} have considered simple vector meson dominance (SVMD)
model in the Glauber approach to describe the nuclear shadowing seen in the
data in the region $1-3$ GeV. In SVMD model, the low-lying vector mesons,
i.e., $V$=$\rho^0$, $\omega$ and $\phi$ mesons, are considered because the
low energy photon beam, as mentioned above, was used in the measurements.
According
to them, the nuclear shadowing can be understood by the proper choice of
$\alpha_{\rho N}$, i.e., the ratio of the real to imaginary part of the
$\rho$-nucleon scattering amplitude.
The
use of $\alpha_{\rho N}$ evaluated by Kondratyuk et al., \cite{kondratyuk}
gives good description of the shadowing in the quoted reaction \cite{falter}.

As pointed out in Ref.~\cite{pautz}, the shadowing in the $\rho$ meson
photoproduction reaction in the multi-GeV region is better accounted by
using generalized vector meson dominance (GVMD) model. In this model, the
higher vector meson states $V^\prime$ are used along with the low lying
vector mesons $V$.
The
elementary scattering of the vector meson in SVMD model is described as
$VN \to VN$, where as the additional processes $V^\prime N \to VN$,
$V^\prime N \to V^\prime N$ and $VN \to V^\prime N$ are incorporated in
GVMD model.
In
fact, this model has been extensively used by many authors for studying the
nuclear shadowing \cite{arneodo}.
For
example, Frankfurt et al., \cite{frank39} show the existence of the hard
and soft components of the virtual photon within GVMD model and predict
the nuclear shadowing in the deep inelastic scattering (DIS) similar to that
within parton model.
Using
Glauber model, the shadowing in the photonuclear reaction is calculated at
$E_\gamma > 3$ GeV where GVMD model is included in the parametric form of
the scattering amplitude used to evaluate the profile function of the
reaction \cite{ditsas}.
The
quoted model is also used to study the scaling behavior of the shadowing
effect in the deep inelastic $\mu$ nucleus scattering \cite{Shaw}.

The shadowing in the photonuclear reaction in the low energy region
($\sim 1-3$ GeV)   is reexamined using both SVMD and GVMD models in Glauber
formalism (modified for the correlated system) for the reaction.
The
photoproduction of the vector meson and its interaction with the nucleus
are described by the measured vector meson nucleon scattering parameters.
The
calculated results for the nuclear shadowing due to SVMD and GVMD models
are compared with the measured spectra in the considered energy region.

\section{Formalism}

The total scattering cross section of the photonuclear reaction
$\sigma_t^{\gamma A}$, as discussed earlier, is composed of the unshadowed
and shadowed parts, i.e.,
\begin{equation}
\sigma_t^{\gamma A}
= A\sigma_t^{\gamma N} + \sigma_{t,V}^{\gamma A},
\label{gAX}
\end{equation}
where $\sigma_t^{\gamma N}$ is the total cross section of the photonucleon
reaction. The first part of the equation, i.e., $A\sigma_t^{\gamma N}$, is
the unshadowed total cross section of the photonuclear reaction, addressed
in Fig.~\ref{FgVp}(a).
The
cross section $\sigma_{t,V}^{\gamma A}$ originates due to the vector meson
scattering on the nucleus (shown in Fig.~\ref{FgVp}(b)), which leads to the
shadowing in the photonuclear reaction. Using the fixed scatterer (or frozen
nucleon) approximation, $\sigma_{t,V}^{\gamma A}$
in SVMD model is given by \cite{yennie}
\begin{eqnarray}
\sigma_{t,V}^{\gamma A} &=& \sum_{V =\rho, \omega, \phi}
                            \frac{8\pi^2}{k_\gamma k_V}
Im \{ i f_{VN \to \gamma N} f_{\gamma N \to VN}
   \int d{\bf b} \int^{+\infty}_{-\infty} dz  \int^{+\infty}_z dz^\prime
\nonumber \\
&~& \times \varrho_2({\bf b}, z^\prime, z) e^{-iq_V (z^\prime -z)} 
  exp \left [ -\frac{1}{2} \sigma_t^{VN} (1-i\alpha_{VN})
      \int^{z^\prime}_z dz_\iota \varrho ({\bf b}, z_\iota)  \right ] \},
\label{vAX}
\end{eqnarray}
where $k_\gamma$ and $k_V$ are the momenta of the incoming $\gamma$ boson
and vector meson respectively. $q_V (=k_\gamma -k_V)$ is the
momentum transfer to the nucleus. All other quantities appearing in the
above equation are described afterwards.
The
widths of the vector mesons are neglected in this equation
\cite{falter, pautz}. It is shown latter that the distinctly dominant
contribution to the nuclear shadowing arises due to the $\rho$ meson.
The
intrinsic width of this meson is so large ($\Gamma_\rho \sim 150$ MeV
\cite{olive}) that it dominantly decays ($\rho^0 \to \pi^+ \pi^-$) inside
the nucleus \cite{das13}. The effect of the $\rho$ meson width is largely
cancelled by the corrections arising from the pions scattering in the
nucleus \cite{pautz}.
The
widths of the $\omega$ and $\phi$ mesons (i.e., $\Gamma_\omega \sim 8.5$
MeV and $\Gamma_\omega \sim 4.3$ MeV \cite{olive, muhli}) is much smaller
than that of the $\rho$ meson, and therefore, they decay outside the nucleus.
The
contributions of the $\omega$ and $\phi$ mesons to the nuclear shadowing,
as discussed latter, are insignificant.
It
should be mentioned that Eq.~(\ref{vAX}) can be interpreted in terms of the
vector meson properties in the nucleus, see the detail in
Ref.~\cite{falter}.

The symbols $\varrho({\bf b}, z_\iota)$ and $\varrho_2({\bf b}, z^\prime, z)$ in
Eq.~(\ref{vAX}) represent one-body and correlated two-body density
distributions of the nucleus.
For
the uncorrelated system (i.e., independent particle approximation),
$ \varrho_2 ({\bf b}, z, z^\prime) $ is expressed in terms of one-body
nuclear densities, i.e.,
$ \varrho_2({\bf b}, z, z^\prime) =
\varrho({\bf b}, z^\prime) \varrho({\bf b}, z) $,
and the cross section in Eq.~(\ref{vAX}) represents the optical model
approximation in Glauber multiple scattering approach in the photonuclear
reaction \cite{bauer, boffi}.
Two-body
correlated nuclear density distribution $ \varrho_2({\bf b}, z^\prime, z) $
is given in Refs.~\cite{falter, boffi}, i.e.,
\begin{equation}
\varrho_2({\bf b}, z^\prime, z)
= \varrho({\bf b}, z^\prime) \varrho({\bf b}, z)
+ \Delta ({\bf b}, |z^\prime -z|),
\label{den2}
\end{equation}
where
$\varrho ({\bf r})$ denotes single particle nuclear density distribution.
The form of it is discussed latter. $ \Delta ({\bf b}, |z^\prime -z|) $
denotes two-body correlation function. Considering Bessel function
parameterization, it can be written as
$\Delta ({\bf b} |z^\prime -z|) =
-j_0 (q_c |z^\prime -z|) \varrho({\bf b}, z^\prime) \varrho({\bf b}, z);$
with $q_c=0.78$ GeV \cite{falter, boffi}. Since
$\varrho_2({\bf b}, z^\prime, z)$ is equal to zero for $z^\prime=z$, there
cannot be overlap of nucleons in the nuclear density distribution.
Therefore, the inclusion of $ \Delta ({\bf b} |z^\prime -z|) $ in nuclear
density distribution avoids the unphysical results.

In Eq.~(\ref{vAX}), $\sigma_t^{VN}$ is the vector meson nucleon total
scattering cross section and $\alpha_{VN}$ denotes the ratio of the real to
imaginary part of the vector meson nucleon scattering amplitude
$f_{VN \to VN}$.
According to simple vector meson dominance (SVMD) model, $f_{VN \to VN}$ is
related to the photoproduction amplitude $f_{\gamma N \to VN}$
\cite{pautz, das13} as 
\begin{equation}
f_{\gamma N \to VN} = f_{VN \to \gamma N}
= \frac{ \sqrt{\pi \alpha_{em}} }{ \gamma_{\gamma V} } f_{VN \to VN},
\label{sca}
\end{equation}
where $\alpha_{em} (=1/137.036)$ is the fine structure constant.
$\gamma_{\gamma V}$ is the photon $\gamma$ to vector meson $V$ coupling
constant which is determined from the measured $V \to e^+e^-$
decay width \cite{olive}:
$ \gamma_{\gamma \rho^0} / \gamma_{\gamma \omega} / \gamma_{\gamma \phi}
= 2.48 / 8.53 / 6.72 $ \cite{sibir06}.

\section{Result and Discussions}

The transparency $T$, i.e.,
$\frac{ \sigma_t^{\gamma A} }{ A\sigma_t^{\gamma N} }$, is calculated
using Eq.~(\ref{gAX}) to study the shadowing in the photonuclear reaction
in the energy region of $1-3$ GeV. The data of the quoted transparency
$T$ exist for  $^{12}$C, $^{27}$Al, $^{63}$Cu, $^{112}$Sn and $^{208}$Pb
nuclei.
The
total cross section $\sigma_t^{\gamma N}$ of the photonucleon reaction is
defined as
$ \sigma_t^{\gamma N} = \frac{Z}{A}\sigma_t^{\gamma p}
                    + \frac{A-Z}{A}\sigma_t^{\gamma n} $,
where $\sigma_t^{\gamma p}$ and $\sigma_t^{\gamma n}$ are the total cross
sections of the gamma-proton $(\gamma p)$ and gamma-neutron $(\gamma n)$ 
scattering respectively. The energy dependent measured values for them
are taken from Refs.~\cite{olive, armst}.

The form of the single particle density distribution of the nucleus
$\varrho ({\bf r})$, as extracted from the electron scattering experiment,
is given in Ref.~\cite{andt}. $\varrho ({\bf r})$ for $^{12}$C is
described by the harmonic oscillator Gaussian form \cite{andt}, i.e.,
\begin{equation}
\varrho (r) = \varrho (0) [ 1 + w (r/c)^2 ] e^{-(r/c)^2},
\label{hod}
\end{equation}
with $w=1.247$ and $c=1.649$ fm.
For
other nuclei quoted above, $\varrho ({\bf r})$ is described by 2pF Fermi
distribution function \cite{andt}:
\begin{equation}
\varrho (r) = \varrho (0) \frac{1}{1+e^{-(r-c)/z}}.
\label{fmd}
\end{equation}
The parameters $c$ (half-density radius) and $z$ (diffuseness) for nuclei
appearing in the equation are taken from Ref.~\cite{andt}. The density is
normalized to the mass number of the nucleus.

The $\rho N$ scattering amplitude $f_{\rho N \to \rho N}$ is taken from
the analysis by Kondratyuk et al., \cite{kondratyuk}. The experimentally
determined imaginary part of $\omega N$ scattering amplitude
$f_{\omega N \to \omega N}$ is given in \cite{sibir, lyka}.
For
the real part of $f_{\omega N \to \omega N}$, the ratio $ \alpha_{\omega N}
( = \frac{ Re f_{\omega N \to \omega N} }{ Im f_{\omega N \to \omega N} } ) $
has been calculated using additive quark model and Regge theory
\cite{sibir}:
$ \alpha_{\omega N}
= \frac{0.173(s/s_0)^\epsilon - 2.726(s/s_0)^\eta}
       {1.359(s/s_0)^\epsilon + 3.164(s/s_0)^\eta} $,
with $s_0=1$ GeV$^2$, $\epsilon = 0.08$ and $\eta = -0.45$.
$s$ is the $\omega N$ cm energy.
According
to vector meson dominance (VDM) model \cite{das13, sibir06}, the forward
cross section of the $\gamma p \to \phi p$ reaction, i.e.,
$\frac{d\sigma}{dq^2} (\gamma p \to \phi p) |_{q^2=0}$, is given by
\begin{equation}
\frac{d\sigma}{dq^2} (\gamma p \to \phi p) |_{q^2=0}
= \frac{\alpha_{em}}{16\gamma^2_{\gamma \phi}}
  \left ( \frac{ \tilde{k}_\phi }{ \tilde{k}_\gamma } \right )^2
  \ [1+\alpha^2_{\phi N}] (\sigma^{\phi N}_t)^2,
\label{ftD2}
\end{equation}
where $\tilde{k}_\phi$ and $\tilde{k}_\gamma$ are the c.m. momenta in the
$\phi N$ and $\gamma N$ systems respectively, evaluated at the c.m. energy
of $\gamma N$ system.
Using
$\alpha_{\phi N} =-0.3$ \cite{bauer} and the parameterized form of
$ \frac{d\sigma}{dq^2} (\gamma p \to \phi p) $ given in Eq.~(3.85a) in
Ref.~\cite{bauer}, the imaginary part of the $\phi N$ scattering amplitude
$f_{\phi N \to \phi N}$ is extracted from the above equation.

\begin{figure}[h]
\begin{center}
\centerline {\vbox {
\psfig{figure=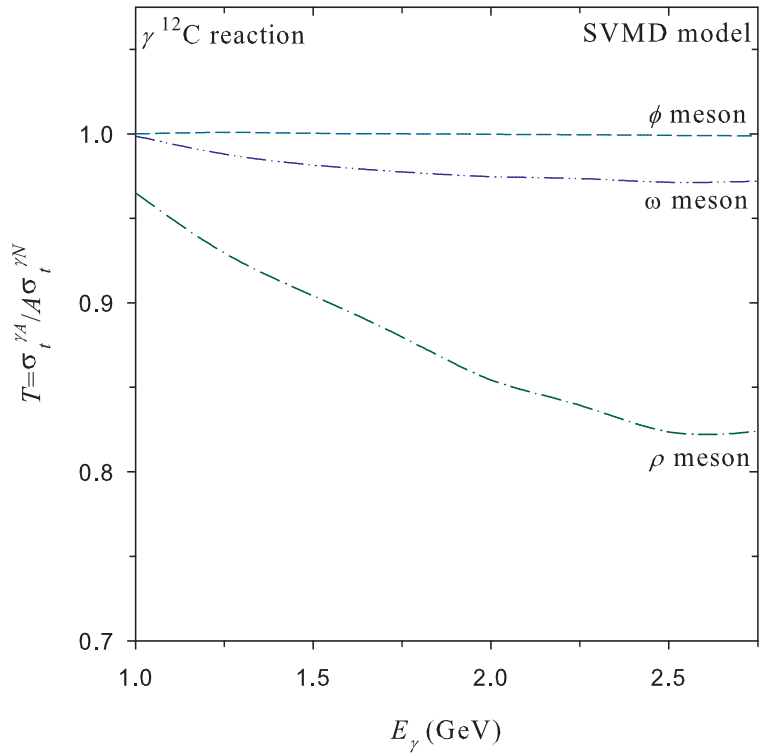,height=10.0 cm,width=10.0 cm}
}}
\caption{
(color online). The shadowing ($T<1$) in the $\gamma ^{12}\mbox{C}$
reaction due to $\rho^0$, $\omega$ and $\phi$ mesons used in SVMD model.
The figure shows that the nuclear shadowing distinctly occurs
due to the $\rho$ meson.
}
\label{FgCm}
\end{center}
\end{figure}

The shadowing in the photonuclear reaction is defined earlier by the
transparency $T =\frac{ \sigma_t^{\gamma A} }{ A\sigma_t^{\gamma N} } <1$.
Amongst the $\rho$, $\omega$ and $\phi$ mesons in SVMD model, the shadowing
in the photonuclear reaction, as shown by the dot-dashed curve in
Fig.~\ref{FgCm}, distinctly originates due to the $\rho$ meson.
The
momentum transfer to the nucleus is large for the massive vector meson
production (see Eq.~\ref{vAX}), which leads to less cross section or
shadowing in the reaction.
Though
the masses $m_\rho$(=775.26 MeV) $\sim$ $m_\omega$(=782.65 MeV) and the
elementary cross sections $\sigma_t^{\omega N} \sim \sigma_t^{\rho N}$,
the $\omega$ meson weakly contributes to the nuclear showing, see the
dot-dot-dashed curve in the figure. It occurs since the $\omega$ meson
photoproduction amplitude, according to Eq.~(\ref{sca}), is the least, i.e.,
$\gamma_{\gamma \omega} = 8.53$.
The
small $\phi$ meson photoproduction amplitude ($\gamma_{\gamma \phi} =6.72$),
and relatively large mass ($m_\phi \sim$ 1020 MeV) leads to 
negligible contribution of the $\phi$ meson (shown by the small dashed
curve) to the nuclear shadowing.

Since the distinctly dominant contribution to the nuclear shadowing (as
illustrated in Fig.~\ref{FgCm}) arises because of the $\rho$ meson, the
higher states of this meson (discussed in GVMD model) are to be considered
as the shadowing due to the $\rho$ meson can be modified because of those
states.
There
exist few higher $\rho$ meson states \cite{olive} but, unfortunately, the
dielectron decay width $\Gamma_{V\to e^+e^-}$ of these mesons are not known.
The
measured $\Gamma_{V\to e^+e^-}$, as mentioned below Eq.~(\ref{sca}), is
used to extract the photon to vector meson coupling constant
$\gamma_{\gamma V}$.
Unless
those (i.e., $\gamma_{\gamma V}$s) are known for the higher $\rho$ meson
states, they cannot be incorporated in the GVMD model to describe 
photon induced reactions.
As
done by Pautz and Shaw \cite{pautz}, the contributions of all higher $\rho$
meson states are approximated with that of an effective state
(called $\rho^\prime$ meson) whose mass and coupling (to photon) constant
can be expressed by those of the $\rho$ meson.
The
calculated results due to them account very good the measured $\rho$ meson
photoproduction data in the multi-GeV region.
Therefore,
the quoted effective $\rho$ meson state, i.e., $\rho^\prime$ meson, is
considered in GVMD model, and the scattering amplitude of the $\rho$ meson,
predicted by SVMD model in Eq.~(\ref{sca}), is replaced by those of the
$\rho$ and $\rho^\prime$ mesons in GVMD model \cite{frank01} as
\begin{eqnarray}
&f_{\gamma N \to \rho N}& =
 \frac{ \sqrt{\pi \alpha_{em}} }{ \gamma_{\gamma \rho} }
  f_{\rho N \to \rho N}
+\frac{ \sqrt{\pi \alpha_{em}} }{ \gamma_{\gamma \rho^\prime} }
f_{\rho^\prime N \to \rho N},     \nonumber  \\
&f_{\gamma N \to \rho^\prime N}& =
 \frac{ \sqrt{\pi \alpha_{em}} }{ \gamma_{\gamma \rho^\prime} }
  f_{\rho^\prime N \to \rho^\prime N}
+\frac{ \sqrt{\pi \alpha_{em}} }{ \gamma_{\gamma \rho} }
f_{\rho N \to \rho^\prime N},     
\label{sag}
\end{eqnarray}
where
$\gamma_{\gamma \rho^\prime}$ denotes the photon to $\rho^\prime$ meson
coupling constant. The properties of the $\rho^\prime$ meson
(i.e., mass $m_{\rho^\prime}$ and $\gamma_{\gamma \rho^\prime}$) are given
by
$m_{\rho^\prime} = \sqrt{3} m_\rho, 
\gamma_{\gamma \rho^\prime}
= \frac{m_{\rho^\prime}}{m_\rho} \gamma_{\gamma \rho} $
\cite{pautz, frank01}.
The scattering amplitudes are related to each other as
$f_{\rho^\prime N \to \rho^\prime N} = f_{\rho N \to \rho N}, 
f_{\rho^\prime N \to \rho N}
= f_{\rho N \to \rho^\prime N} = -\epsilon f_{\rho N \to \rho N},$
with $\epsilon =0.18$ \cite{pautz, frank01}. The above equations illustrate
the reduction of $f_{\gamma N \to \rho N}$ due to the inclusion of
$\rho^\prime$ meson (GVMD model) which leads to less shadowing in the
photonuclear reaction.
As
explained earlier, the nuclear shadowing due to the vector meson decreases
with the increase in its mass $m_V$ and coupling constant to photon
$\gamma_{\gamma V}$.
Due
to these reasons, the contribution of the higher states of the vector mesons
(i.e., $\rho^\prime$, $\omega^\prime$ and $\phi^\prime$ mesons) to the
quoted shadowing is very less.
The
$\omega^\prime$ and $\phi^\prime$ mesons are not included in the
respective photoproduction amplitudes $f_{\gamma N \to \omega N}$ and
$f_{\gamma N \to \phi N}$,
since the contribution of the $\omega$ and $\phi$ mesons (SVMD model) to
the shadowing in the considered reaction (as shown in Fig.~\ref{FgCm}) is
insignificant.
The
inclusion of the $\omega^\prime$ and $\phi^\prime$ mesons would further
reduce the nuclear shadowing due to the $\omega$ and $\phi$ mesons
respectively, which are already negligibly small.

To compare the nuclear shadowing in SVMD and GVMD models, the
transparencies $T$ are calculated using those models, and the calculated
results along with the data \cite{falter, bianchi} are presented in
Fig.~\ref{FRAN}.
The
dot-dashed curves describe the nuclear shadowing due to SVMD model, i.e.,
the $\rho$, $\omega$ and $\phi$ mesons are taken into account to describe
the reaction.
The
solid curves arise because of GVMD model, i.e., the effective higher
$\rho$ meson state ($\rho^\prime$ meson), in addition to above 
vector mesons, is used to calculate the nuclear transparency.
Fig.~\ref{FRAN}
distinctly shows that the nuclear shadowing is reduced because of the
$\rho^\prime$ meson in GVMD model, and the bump appearing at
$E_\gamma \sim 1.6$ GeV for heavier nuclei is well reproduced by this model.
The calculated results are well accord with the data of the shadowing in
the photonuclear reaction.

The total cross section per nucleon of the photonuclear reaction, i.e.,
$\sigma_t^{\gamma A}/A$ calculated using Eq.~(\ref{gAX}), are compared
with the data \cite{bianchi} in Fig.~\ref{FgAN}.
The
dot-dashed curves in this figure occur due to SVMD model where as the
solid curves arise because of GVMD model. The figure shows that the cross
section increases because of the $\rho^\prime$ meson in GVMD model.
The calculated results reproduce the data reasonably well.

Figs.~\ref{FRAN} and \ref{FgAN} show that the SVMD model based calculated
results reproduce well the measured spectra for lighter nuclei, i.e.,
$^{12}$C and $^{27}$Al, through out the considered beam energy region.
Those
results for heavier nuclei (i.e., $^{63}$Cu, $^{112}$Sn and $^{208}$Pb)
are also well accord with data except the beam energy region
$E_\gamma \sim 1.5-1.7$ GeV, where they underestimate the measured spectra.
It
is noticeable that the calculated results due to GVMD model work other way.
The results based on this model overestimate the data for the lighter nuclei
in the energy region  $E_\gamma \sim 1.5-1.7$ GeV. Beyond this region,
those results are well accord with the data.
For
the heavier nuclei, the calculated results due to GVMD model reproduce the
measured spectra remarkably well.

\section{Conclusions}

The shadowing of the photonuclear reaction in the $1 -3$ GeV region are
studied using the optical theorem in Glauber approach for the multiple
scattering of the vector mesons in the nucleus.
Both
SVMD and GVMD models are used to interpret the vector meson production
in the quoted reaction. In the previous model, the low-lying vector mesons
(i.e., $\rho^0$, $\omega$, and $\phi$ mesons) are considered,
whereas the later model adds an effective higher $\rho$ meson state,
i.e., $\rho^\prime$ meson.
The
shadowing in the photonuclear reaction distinctly occurs because of the
$\rho$ meson. 
The
cross section of the photonuclear reaction increases due to the inclusion
of the $\rho^\prime$ meson (GVMD model) which leads to the reduction of the
shadowing in the reaction.
The
calculated results are well accord with data. Specifically, those in GVMD
model reproduce noticeably well the bump appearing in the measured spectrum
for the heavy nuclei.

\section{Acknowledgement}

The author thanks anonymous referee for the comments which improve the
quality of the paper. S. H. Gharat is acknowleged for preparing figure~1
in the paper.

\begin{figure}[h]
\begin{center}
\centerline {\vbox {
\psfig{figure=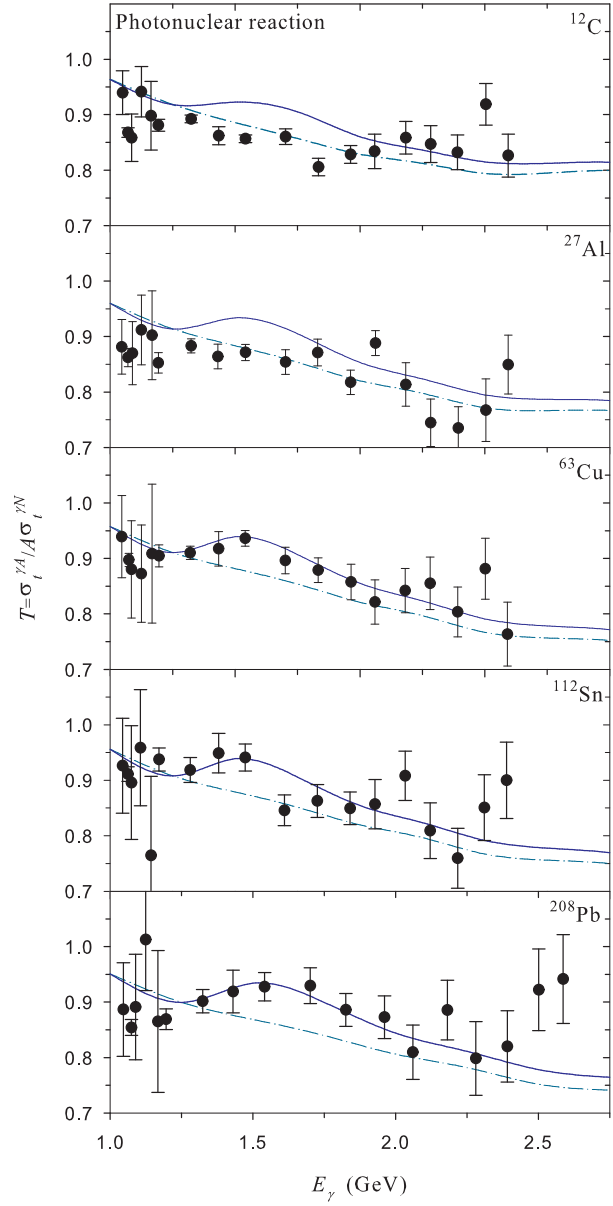,height=16.0 cm,width=08.0 cm}
}}
\caption{
(color online). The shadowing in the photonuclear reactions compared with
data. The dot-dashed curves describe the nuclear transparencies due to SVMD
model, whereas the solid curves represent those due to GVMD model (see text).
}
\label{FRAN}
\end{center}
\end{figure}
\begin{figure}[h]
\begin{center}
\centerline {\vbox {
\psfig{figure=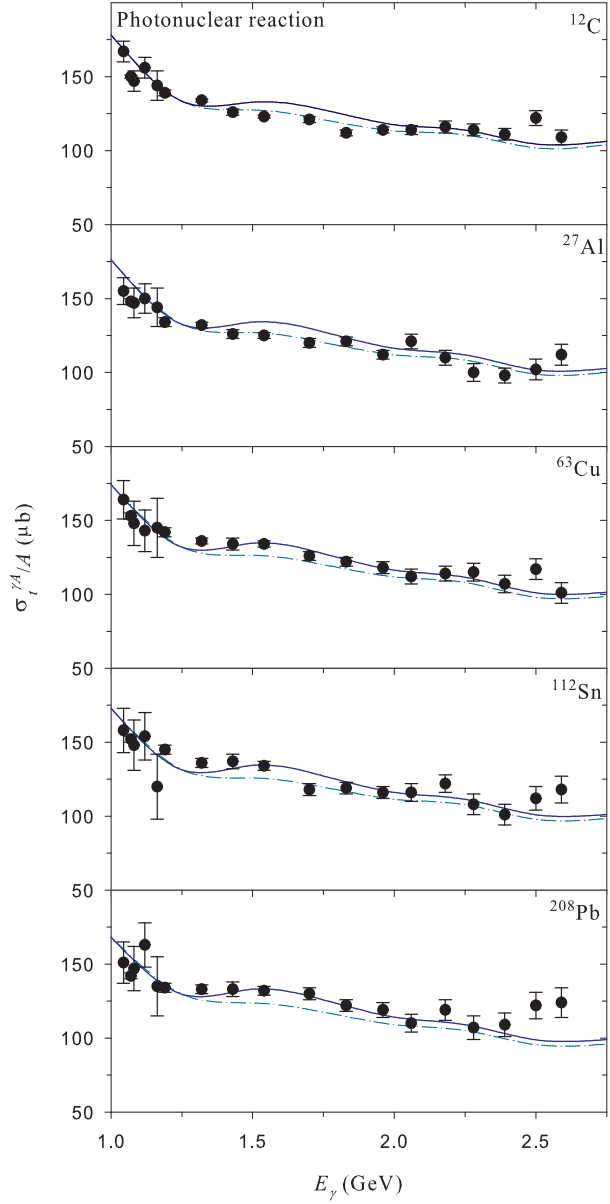,height=16.0 cm,width=08.0 cm}
}}
\caption{
(color online). The total cross section per nucleon of the photonuclear
reactions presented with the measured spectra \cite{bianchi}. The curves
represent the calculated results due to SVMD and GVMD models, as
explained in Fig.~\ref{FRAN}.
}
\label{FgAN}
\end{center}
\end{figure}

{\bf Figure Captions}
\begin{enumerate}

\item
The pictorial presentation of Compton scattering on the nucleus.

\item
(color online). The shadowing ($T<1$) in the $\gamma ^{12}\mbox{C}$
reaction due to $\rho^0$, $\omega$ and $\phi$ mesons used in SVMD model.
The figure shows that the nuclear shadowing distinctly occurs
due to the $\rho$ meson.

\item
(color online). The shadowing in the photonuclear reactions compared with
data. The dot-dashed curves describe the nuclear transparencies due to SVMD
model, whereas the solid curves represent those due to GVMD model (see text).

\item
(color online). The total cross section per nucleon of the photonuclear
reactions presented with the measured spectra \cite{bianchi}. The curves
represent the calculated results due to SVMD and GVMD models, as
explained in Fig.~\ref{FRAN}.

\end{enumerate}

\newpage

\newpage

\newpage

\newpage

\end{document}